\documentclass[12pt]{article}
\usepackage{epsf}
\usepackage{graphicx}
\usepackage{lscape}
\usepackage{amssymb}
\usepackage{pazh}
\tightenlines

\def\ΧΆ{$\pm$}
\def\*{$^{*}$}

\def\a{$^{\mbox{\small a}}$}
\def\b{$^{\mbox{\small b}}$}
\def\c{$^{\mbox{\small c}}$}
\def\d{$^{\mbox{\small d}}$}
\def\e{$^{\mbox{\small e}}$}

\def\Œ³ŒÀ{\mbox{\<<Œ³ŒàŒÐŒÝŒÐŒâ\>>}}

\def\ŒÕŒàŒÓŒá{ŒíŒàŒÓ~Œá$^{-1}$}
\def\ŒÕŒàŒÓŒáŒÜ{ŒíŒàŒÓ~ŒáŒÜ$^{-2}$~Œá$^{-1}$}
\def\etal{{\it et~al.}}

\begin{document}
 
{\footnotesize Astronomy Letters, Vol. 31, No. 2, 2005, pp. 88-97. 
Translated from Pis'ma v Astronomicheskii Zhurnal, Vol. 31, No. 2, 
2005, pp. 99-109.
Original Russian Text Copyright \copyright\, 2005 by Tsygankov, 
Lutovinov.}


\title{\bf Observations of the Transient X-ray Pulsar KS 1947+300 
by the INTEGRAL and RXTE Observatories}   
 
\author{\bf \hspace{-1.3cm}\copyright\, 2005 \ \
   S.S.Tsygankov\affilmark{1}$^{\,*}$, A.A.Lutovinov\affilmark{1}}     
 
\affil{
$^1$ {\it Space Research Institute, Russian Academy of Sciences, 
Profsoyuznaya ul. 84/32, Moscow 117810, Russia}}
 
\vspace{2mm}

\sloppypar 
\vspace{2mm}
\noindent


{  
We analyze the observations of the X-ray pulsar KS 1947+300 
performed by the \mbox{INTEGRAL}
and RXTE observatories over a wide (3-100 keV) X-ray 
energy range. The shape of the pulse profile was
found to depend on the luminosity of the source. Based 
on the model of a magnetized neutron star, we
study the characteristics of the pulsar using the change 
in its spin-up rate. We estimated the magnetic field
strength of the pulsar and the distance to the binary. 
\copyright\, 2005 Pleiades Publishing Inc.}


\vfill
 
{$^{*}$ E-mail: st@hea.iki.rssi.ru}
\newpage
\thispagestyle{empty}
\setcounter{page}{1}

\section*{INTRODUCTION}

    The transient X-ray source KS 1947+300 was
discovered in June 1989 during the observations of
the Cyg X-1 region by the TTM telescope aboard
the Kvant module of the Mir space station (Borozdin
\etal~ 1990). The flux recorded from it was $70\pm±
10$ mCrab in the energy range 2-27 keV, and the
spectrum was fitted by a power law with an index
of $\gamma=-1.72\pm0.31$ and low-energy photoabsorption
with an atomic hydrogen column density of $N_H =
(3.4\pm3.0) \times 10^{22}$ cm$^{-2}$. In April 1994, the BATSE
monitor of the ComptonGRO observatory discovered
the X-ray pulsar GRO J1948+32 with a period of
18.7 s in the same region of the sky. The spectrum
of this source in the energy range 20-120 keV was
described by a power law with a photon index of $\gamma=
2.65\pm0.15$ (Chakrabarty \etal~ 1995). Subsequently,
KS 1947+300 and GRO J1948+32 were found to be
the same object, a transient X-ray pulsar (Swank and
Morgan 2000).

    Until recently, no abrupt changes in pulsation period 
(the so-called glitches) were known in accreting
X-ray pulsars. However, having analyzed the RXTE
data, Galloway \etal~ (2004) pointed out that the
pulsation period of KS 1947+300 changes over a very
short time. Thus, in January 2001, its pulsation frequency 
increased by about $1.8\times10^{-6}$~ Hz in less than
10 h (the rate of change is $\dot
f \approx 5 \times 10^{-11}$ Hz s$^{-1}$),
while the X-ray flux exhibited no significant increase.
The fractional change in frequency in this case was
$\frac{\Delta\nu}{\nu}=3.7\times10^{-5}$, 
which is much larger than that
observed during glitches in radio pulsars (Krawczyk
\etal~ 2003) and anomalous X-ray pulsars (Kaspi and
Gavriil 2003). At the same time, a comparison with
the BATSE data reveals a spin-down of the pulsar
at a mean rate of $\dot f \approx -8 \times 10^{-13}$~
Hz s$^{-1}$~ on a time
scale of several years. Based on the behavior of the
pulsation period during the outburst of 2000-2001,
Galloway \etal~ (2004) determined the parameters of
the binary: the orbital period $P_{orb}=40.415\pm0.010$~ d,
the projected semimajor axis of the relativistic object
$a_{x}$sin$i=137\pm3$~ light seconds, and the eccentricity
$e=0.033\pm0.013$.
     The optical counterpart in the binary is a B0 Ve
star. If the true luminosity of the star is typical of
stars of this spectral type, then the distance to it is
estimated as $\sim10$~ kpc (Negueruela \etal~ 2003).
     In this paper, we present the results of our spectral 
and timing analyses of the pulsar KS1947+300
based on INTEGRAL and RXTE data in a wide (3-­
100 keV) energy range.

\section*{OBSERVATIONS}

     The INTEGRAL International observatory (Winkler 
\etal~ 2003) placed in orbit by a Russian Proton
launcher on October 17, 2002, carries four scientific
instruments that allow the emission from astrophysical 
objects to be studied over a wide wavelength range
(from optical to hard $\gamma$-rays). Here, we used mainly
data from the ISGRI detector of the IBIS gamma-
ray telescope (Ubertini \etal~ 2003), which includes
two layers of position-sensitive detectors (ISGRI and
PICsIT) and a coded mask. The ISGRI detector is
effectively sensitive to photons in the energy range 20
to 200 keV (the energy resolution is $\sim$7\% at 100 keV)
and can image the sky region within a 
29$^{\circ}\times$29$^{\circ}$ field
of view (the full coding zone is 9$^{\circ}\times$9$^{\circ}$) 
with a nominal
spatial resolution of $\sim$12 arcmin (the angular size of
the mask element). See Lebrun \etal~ (2003) for a
more detailed description of the detector.

   The X-ray pulsar KS 1947+300 occasionally falls
within the field of view of the scientific instruments
when the Galactic plane is scanned as part of the
INTEGRAL Main Observing Program. In this 
paper, we analyze the INTEGRAL observations 
performed from December 2002 through April 2004.
About 700 pointings were made over this period in
which the object under study fell within the field of
view of the INTEGRAL instruments; the total exposure 
time was $\sim1.5\times10^6$ s.

   We performed the image reconstruction and the
spectral analysis of the IBIS data using the methods
described by Revnivtsev \etal~ (2004) and Lutovinov
\etal~ (2003). An analysis of the large set of 
calibration observations for the Crab Nebula revealed a
systematic error of $\sim$10\% in the measured absolute
flux from the source over a wide energy range; the
spectral shape was reconstructed with an accuracy
as high as 3-5\%. This was taken into account in
the spectral analysis by adding a systematic error of
5\%. All of the errors given in this paper are purely
statistical and correspond to one standard deviation.
The standard OSA software of version 3.0 was used
for our timing analysis of the source on time scales
of the pulsation period and for our spectral analysis of
the JEM-X data.

   Figure 1 shows the sky map with the X-ray pulsar
KS 1947+300 obtained by the IBIS telescope in the
energy range 18-60 keV. This map was constructed
when the Galactic plane was scanned on May 11,
2003 (MJD 52770). The detection significance of the
source was 29 $\sigma$ at a total exposure of  $\sim$6 ks.

    Since the object under study was not within the
field of view of the JEM-X X-ray monitor (Lund
\etal~ 2003) aboard the INTEGRAL observatory in
the overwhelming majority of the cases, we failed
to analyze its behavior in the standard X-ray energy 
range. Therefore, we used simultaneous data
from the ASM monitor of the RXTE observatory
(http://xte.mit.edu/ASM$_{-}$lc.html) to make up for
the lack of information in this range (1.3-12.2 keV).
In addition, based on publicly available RXTE data,
we traced the evolution of the emission from the
source over a wide energy range during one of its
previous outbursts in 2000-2001 (Obs. ID 50068,
50425, and 60402). The main instruments of the
RXTE observatory (Bradt \etal~ 1993) are the PCA
and HEXTE spectrometers with the operating energy
ranges 3-20 and 15-250 keV, respectively. The
PCA spectrometer is a system of five xenon/propane
proportional counters with an effective area of 
$\sim$6400 cm$^2$ at 6-7 keV; its energy resolution at
these energies is $\sim$18\%. The HEXTE spectrometer 
is a system of two independent packages of
NaI(Tl)/CsI(Na) phoswich detectors swinging with
an interval of 16 s for the observations of background
areas at a distance of 1.5$^\circ$ from the source. At each
specific time, the source can be observed only by one
of the two detector packages; thus, the effective area
of the HEXTE detectors is $\sim$700 cm$^2$.

    The standard FTOOLS/LHEASOFT 5.3 software 
package was used to process the RXTE data.

\section*{TIMING ANALYSIS}

    As was mentioned above, the source KS 1947+300
exhibits strong outburst activity. Figure 2 shows its
light curve in the energy range 18-60 keV 
constructed from the IBIS/INTEGRAL data in 2002-
2004 (Fig. 2a) and the corresponding light curve in
the energy range 1.3-12.2 keV constructed from the
ASM/RXTE data (Fig. 2b).

    As we see from the light curve, the source was in a
low state during the first series of INTEGRAL 
observations in December 2002 (MJD 52600-52650); its
intensity was $\sim$4.5 mCrab in the energy range 18-
60 keV. At the end of this period, the flux from the
source began to increase and reached $\sim$30 mCrab,
but we failed to observe the outburst in full.

    During the subsequent observations, the pulsar
KS 1947+300 was within the IBIS field of view
much more rarely, which, however, allowed several 
intense outbursts to be detected from it. The
first began in mid-April 2003 (MJD 52750) and
lasted about 50 days, and the peak flux from the
source was about 80 mCrab in the energy range
18-60 keV. Subsequently, we detected two more
outbursts from this object in December 2003 and
April 2004 with peak fluxes of $\sim$70 (MJD 52985)
and $\sim$100 (MJD 53102) mCrab in the energy range
18-60 keV, respectively. Figure 2 shows a clear
correlation between the hard and soft X-ray fluxes
from the source.

    Table 1 gives the fluxes from the pulsar
KS 1947+300 measured in the energy range 18-
60 keV, the orbital phases during the observations 
under consideration calculated using $T_{\pi/2}=51985.31$~MJD
from Galloway \etal~ (2004), and its
pulsation periods determined by an epoch-folding
technique after the photon arrival times were 
corrected for the motion of the Earth, the spacecraft,
and the neutron star in the binary. As we see from
Table. 1, the pulsation frequency of the source near
the peak of the April 2003 (MJD 52760-52780)
outburst is proportional to the flux from it. Since
the observational data and the statistics are scarce
(see, e.g., the measurements of the pulsation period
during the December 2003 outburst), the significance
of this result is low. However, it agrees with the
ComptonGRO and RXTE measurements of the
pulsation period during previous outbursts.

    An analysis of the emission from X-ray pulsars
indicates that their pulse profiles can strongly depend
on the energy and intensity of the source (see,
e.g., White \etal~ 1983; Nagase 1989; Lutovinov \etal~
 1994; and references therein). We studied the
behavior of the pulse profile and the pulse fraction
for KS 1947+300 as a function of its state. Figure 3
shows the phase light curves for the pulsar obtained
from the IBIS/INTEGRAL data and averaged at
different intensities: the flux from the source in the
energy range 18-60 keV is (a) $F\approx78$~mCrab
(MJD 52770), (b) $F\approx48$~mCrab (MJD 52994), and
(c) $F\approx5$~mCrab (MJD 526005-52615). In the first
case, the pulse profile is a single broad peak whose
intensity decreases insignificantly with increasing
phase. A finer profile structure (separation into several
individual peaks) begins to show up as the intensity
of the source decreases. During the observations on
April 7, 2004 (MJD 53102), when the peak flux from
the source in the energy range 18-60 keV was about
96 mCrab, the pulse profile had a shape identical to
that shown in Fig. 3a.

    For the subsequent analysis of the results 
obtained and their comparison with theoretical models,
we should pass from fluxes to luminosities over a
wide energy range. However, in most cases, this is
difficult to do, because we have only hard X-ray data
at our disposal. Therefore, to roughly estimate the
bolometric luminosity of the pulsar during the 
INTEGRAL observations of the April-May 2003 outburst,
we used the following method: assuming the main
energy release to be in the energy range 2-100 keV
and the distance to the object to be $d=10$ kpc, we
determined its bolometric luminosity during the 
observations on April 7, 2004, when the source under
study was also within the JEM-X field of view. The
luminosity of the object can then be estimated from
simple proportionality considerations by comparing
the IBIS hard X-ray fluxes at this point and the point
of interest. However, it should be understood that this
estimate is valid only if the spectral shape is constant
at the two points being compared. Thus, the pulse
profiles shown in Fig. 3 correspond to the following
approximate bolometric luminosities of the source: (a)
$2.5\times10^{37}$, (b) $1.5\times10^{37}$, and 
(c) $0.2\times10^{37}$ erg s$^{-1}$, respectively.

    For comparison, Fig. 4 shows the HEXTE/RXTE
pulse profiles for KS 1947+300 in the energy range
18-60 keV. The observations were performed during
the outburst of the source under study that began
in December 2000. Each panel of this figure 
corresponds to the source's mean bolometric luminosities
of $10.6\times10^{37}$ (a), $5.4\times10^{37}$ (b), 
$3.4\times10^{37}$ (c), $0.9\times10^{37}$ (d), 
$0.3\times10^{37}$ (e) erg s$^{-1}$. A significant
difference between this outburst and the outbursts
detected by the INTEGRAL observatory is a factor
of $\sim$3 longer duration and a factor of $\sim$4 higher 
intensity. Nevertheless, the behavior of the pulse 
profiles in this case is similar to that observed by 
INTEGRAL (Fig. 3). In the brightest state, the pulse
profile is a single broad peak with its separation into
two subpeaks at the vertex, one of which is much
narrower than the other; the intensity of the profile
decreases with increasing phase. As the luminosity of
the source decreases, the separation into several 
subpeaks becomes increasingly distinct. When the 
luminosity of the object reaches $0.9\times10^{37}$ erg 
s$^{-1}$, the
profile again becomes double peaked, with the mean
peak shifting backward by approximately a quarter
of the phase. This behavior of the profile may be 
attributable to different emission regimes, depending on
the source's luminosity (Basko and Sunyaev 1976).
The PCA/RXTE data in the softer energy range also
reveal such a dependence of the pulse shape on the
source's intensity, but it is not so distinct (see Fig. 3
from Galloway \etal~ 2004).

Figure 5 shows how the pulse profile for the object
under study changes with energy range. Figures 5a
and 5b present the phase light curves obtained from
the PCA/RXTE observations on February 10, 2001
        (MJD 51950); Fig. 5c shows the same light curves
        obtained from the HEXTE data averaged over the
        period during which the source was near maximum
      light (MJD 51941-51959). The relative intensity of
   the first peak increases with energy, but no significant
         changes in the shape were found.

Since the background cannot be properly determined 
at this time, we were unable to analyze the 
behavior of the pulse fraction using the
    IBIS/INTEGRAL data. Therefore, we used data
     from the HEXTE and PCA spectrometers of the
    RXTE observatory for such an analysis. In Fig. 6a,
   the pulse fraction, which is defined as 
$P=(I_{max}-I_{min})/(I_{max}+I_{min})$, where
$I_{max}$ and $I_{min}$ are the
   background-corrected count rates at the maximum
   and minimum of the pulse profile, is plotted against
    orbital phase; this dependence was obtained from the
HEXTE data in the energy range 18-60 keV near the
peak of the 2000-2001 outburst. We see that there
is a minimum near a phase of $\sim$0.5. A similar, but
slightly less distinct dependence is also typical of the
PCA data in the energy range 3-20 keV (Fig. 6b).
The small scatter of data points in pulse fraction at
close orbit phases is attributable to the dependence of
this parameter on the source's intensity.

\section*{SPECTRAL ANALYSIS}

 As we showed in the previous section, the source
KS 1947+300 exhibits a luminosity dependence of
the pulse profile during X-ray outbursts. Therefore, it
is of particular interest to study the spectral behavior
of the object as a function of the outburst phase.
   
In the standard X-ray energy range, the pulsar
KS 1947+300 was recorded at a statistically significant 
level by the JEM-X telescope of the INTEGRAL
observatory only once, during the observations on
April 7, 2004, when its bolometric luminosity was
$L_{x}\simeq3.1\times
10^{37}$ erg s$^{-1}$. The spectrum of the source
over a wide energy range is well described by a typical
(for this class of objects) model that includes a simple
power law with an exponential cutoff at high energies:

\begin{equation}
I(E)=A\,E^{-\alpha} \times  
\left\{\begin{array}{c} 1, 
\mbox{ ÅÓÌÉ $E<E_{c}$};\\ 
\exp{[-(E-E_{c})/E_{f}\,]}, 
\mbox{ÅÓÌÉ}\ E\geq E_{c},\\
\end{array}\right.
\end{equation} 

where $E$ is the photon energy in keV, $A$ is the 
normalization of the power-law component, $\alpha$ is the photon
spectral index, $E_{c}$ is the cutoff energy, and $E_{f}$ is the 
e-folding energy in the source's spectrum. This model
has long and widely been used to fit the spectra of 
X-ray pulsars (White \etal~ 1983). The energy spectrum
of KS 1947+300 reconstructed from the JEM-X and
IBIS data is shown in Fig. 7, and its best-fit 
parameters are given in Table 2.

 During the remaining INTEGRAL observations,
the pulsar was recorded at a statistically significant 
level only by the ISGRI detector of the IBIS
telescope, and its spectrum could be reconstructed
at energies above 18 keV. When the spectra of
the source in these sessions were described by the
bremsstrahlung model in all three cases (where the
spectrum could be reconstructed), its temperature
remained approximately the same, within the error
limits, and equal to $kT\sim33$ keV. Table 2 gives
the best-fit parameters for the same spectra based
on the model that was used above to describe the
broad-band spectrum, but we fixed the parameters
whose values were outside the ISGRI energy range
at the values that we obtained when analyzing the
spectrum measured on April 7, 2004. We see that the
e-folding energy in the source's spectrum $E_{f}$ (which
is determined in this approach) remains almost constant. 
The relatively large $\chi^{2}$ value for some of the
spectra is attributable to the poor statistics in these
observations.

    We used the RXTE data obtained during the
2000-2001 outburst to analyze the spectral behavior
of the source over a wide energy range in mode detail.
Figure 8 shows the energy spectra averaged over the
same periods as those in which the pulse profiles were
obtained (i.e., February, March, April, and May 2001,
respectively); in June, the luminosity of the source
was too low to reconstruct its spectrum from the
HEXTE data, and only its soft part derived from the
PCA data is shown in the figure. We used the PCA
and HEXTE data for the energy ranges 4-20 and
20-100 keV, respectively. The RXTE data revealed
a feature in the source's spectrum related to the
emission line of neutral iron ($\sim$6.4 keV). The best-fit
parameters for the pulsar's spectra in different states
are given in Table 2. Interestingly, as the intensity of
the source decreases, its spectrum becomes slightly
harder, while the characteristic energies $E_{c}$ and $E_{f}$
decrease (for the observations in June 2001, the fit is
based only on the PCA data). A comparison of the
data from the two observatories shows that the best-fit 
parameters for the INTEGRAL data obtained in
April 2004 also fall on this dependence.

\section*{DISCUSSION}

\subsection*{\it Evolution of the Pulse Profile}

   Basko and Sunyaev (1976) showed the existence
of a critical luminosity $L^*$ ($\sim10^{37}$~erg s$^{-1}$) that 
separates two accretion regimes: the regime in which
the effect of radiation on the falling matter may be
disregarded and the regime in which this effect is
significant. For $L<L^*$, the free-fall zone extends
almost to the neutron-star surface, and the polar cap
radiates mainly upward. In the opposite case ($L>L^*$), 
accretion columns that are elongated along the
magnetic field lines and that radiate predominantly
sideways are formed at the poles.
  
 Based on this model, Parmar \etal~ (1989) first 
explained the luminosity dependence of the pulse profile
for the transient X-ray pulsar EXO 2030+375. These
authors modeled the pulse profile using a simple 
geometric model where the radiation is emitted from the
magnetic poles of a rotating neutron star displaced
from its spin axis. The study was carried out over
a wide luminosity range: from $10^{36}$ to $10^{38}$ 
erg s$^{-1}$.
It was pointed out that the upward-directed 
radiation becomes dominant as the source's luminosity
decreases.

   For the pulsar KS 1947+300, the behavior of its
pulse shape is similar to that described above. For
our analysis, we took the IBIS and HEXTE data
for two different outbursts with luminosity ranges
$(0.2-3.1)\times10^{37}$~ and ~$(0.3-10.6)\times10^{37}$
erg s$^{-1}$, respectively. We consider the 
pulse profiles in the hard
energy ranges due to their relative independence of
external factors, in particular, the weaker dependence
of the shape on the absorption far from the stellar
surface. In both series of observations, the luminosity
passes through its critical value of $L^*$, which is 
reflected in the change of the pulse profile shape (see
Figs. 3 and 4). At high luminosities (i.e., during the
formation of accretion columns), one might expect
the spectrum to be softer than that in the low state,
as confirmed by our spectral analysis.

\newpage
\subsection*{\it The Cyclotron Lines and the Magnetic Field}

   In searching for features in the source's spectrum
related to the resonance cyclotron absorption line, we
added the corresponding component to the model fit
in our spectral analysis. The energy of the line center
$E_{cyc}$ was varied over the range 5-90 keV at a 5-keV 
step, and its width was fixed at 5 keV. Using
the $\Delta\chi^2$ test, we found the most probable position
of the possible cyclotron line at an energy of about
70 keV, but the significance of this feature does not
exceed $\sim2\sigma$. Thus, we can presently conclude that
either the sensitivity and effective exposure time of the
modern INTEGRAL and RXTE instruments are not
enough to detect the cyclotron line in the spectrum of
KS 1947+300 or it lies outside the energy range 5-100 keV. 
If the latter is true, then the surface magnetic
field of theneutron star must be either $<5.6\times10^{11}$
or $>10^{13}$ G.

\subsection*{\it Evolution of the Pulsation Period}

   During outbursts in X-ray pulsars, the rate of 
accretion onto the neutron star increases significantly.
In this case, a spin-up of the pulsar attributable to
the increase in the amount of angular momentum
transferred by the accreted matter to the neutron star
can be observed, with the magnetic field strength of
the neutron star playing an important role.
  A correlation between the rate of change in the
pulse period and the X-ray luminosity during 
outbursts has now been found for seven transient sources
(Galloway \etal 2004; Baykal \etal 2002, and 
references therein). As was noted above, during one of
the outbursts observed by INTEGRAL (April 2003),
there is a direct correlation between the flux from
the source and its pulsation frequency similar to that
observed by RXTE during the 2000-2001 outburst.
    
Based on the observed parameters of the pulsar
KS 1947+300 during its outbursts, we can attempt to
estimate its magnetic field strength and the distance
to the binary using the model of a magnetized neutron
star (Ghosh and Lamb 1979). The following relation
must hold in the case of accretion from the disk:

\begin{equation}\label{2}
\dot \nu \varpropto \mu^{2/7} n(\omega_s) L^{6/7} = \mu^{2/7} n(\omega_s)
(4\pi d^2 F)^{6/7}, 
\end{equation}

where $\mu$µ is the magnetic moment of the neutron star
with a magnetic field $B$ and radius $R$, $n(\omega_s)$ is a
dimensionless function that depends on the fastness
parameter $\omega_s$, $d$ is the distance to the binary, and 
$F$ is the X-ray flux from it.

    In Fig. 9, the rate of change in the pulsation
frequency is plotted against the flux recorded from
KS 1947+300. The circles in the figure correspond
to the measured spin-up and spin-down rates of the
pulsar near the peak of the April-May 2003 outburst,
as inferred from INTEGRAL data, and the spin-up
rate of the pulsar during the outburst of 2000-2001,
as inferred from the RXTE data. When fitting these
data by the function given by Eq. (2), we fixed the
mass and radius of the neutron star at $1.4 M_{\odot}$ and
$10^6$ cm, respectively. As a result, we obtained the
following values: the distance to the source 
$d=9.5\pm1.1$ kpc and the magnetic field of the neutron star 
$B=2.5^{+0.4}_{-1.1} \times 10^{13}$ G, which corresponds 
to the position 
of the cyclotron feature in the object's spectrum at
an energy of about 220 keV. The derived values agree
with the distance to the binary estimated from optical
observations (Negueruela \etal~ 2003) and with the
magnetic field strength estimated by analyzing the
source's spectra (see above). It should be noted that
we used the expression 
$n(\omega_s)=1+\frac{20(1-1.94\omega_s)}{31(1-\omega_s)}$
for the dimensionless angular momentum $n(\omega_s)$ from
the paper by Li and Wang (1996). When the function
suggested by Ghosh and Lamb (1979) is used as a
fit to $n(\omega_s)$, the magnetic field strength proves to be
slightly smaller, $B\sim1.6 \times 10^{13}$ G, but the distance
to the binary increases significantly, $d\sim14$ kpc. If the
latter parameter is fixed at 10 kpc, then the surface
magnetic field of the neutron star decreases to  
$B\sim5\times 10^{12}$ G, but the quality of the fit 
to the data points
in Fig. 9 deteriorates significantly.
 
   As was noted in the Introduction, the mean spin-down 
rate of the pulsar KS 1947+300 is low enough
for its pulsation period to be considered close to
the equilibrium value. In the case of disk accretion,
this period is then related to the parameters of
the neutron star as follows: 
$P_{eq}\simeq1.0\, L_{37}^{-3/7}\mu_{30}^{6/7}$ s
(Lipunov 1987). Assuming that the the source's
luminosity is $\sim10^{37}$ erg s$^{-1}$ and the pulsation period
is $\sim18.7$ s, we obtain a magnetic field strength of
$\sim3\times10^{13}$ G, in good agreement with the above
estimates.

\section*{ACKNOWLEDGMENTS}

   We wish to thank E.M. Churazov, who developed
the methods and software for analyzing the data from
the IBIS telescope of the INTEGRAL observatory.
We also wish to thank M.G. Revnivtsev for help
in processing the RXTE data and valuable remarks
and discussions. This work was supported by the
Ministry of Industry and Sciences (grant no. 
NSh-2083.2003.2 from the President of Russia and project
no. 40.022.1.1.1102) and the Russian Foundation for
Basic Research (project nos. 02-02-17347 and 
04-02-17276). We used the data retrieved from the 
High-Energy Astrophysics Archive at the Goddard Space
Flight Center of NASA and the data retrieved from
the Archive of the INTEGRAL Science Data Center
(Versoix, Switzerland) and the Russian INTEGRAL
Science Data Center (Moscow Russia).

 \pagebreak

\section*{REFERENCES}

1. M. M. Basko and R. A. Sunyaev, MNRAS 175, 395
   (1976).

2. A. Baykal, M. J. Stark, and J. H. Swank, Astrophys.
   J. 569, 903 (2002).

3. K. N. Borozdin, M. R. Gilfanov, R. A. Sunyaev, \etal
   Sov. Astron. Lett. 16, 345 (1990) [Pis'ma Astron. Zh. 16, 804 (1990)]. 

4. H. V. Bradt, R. E. Rothschild, and J. H. Swank,
   Astron. Astrophys. Suppl. Ser. 97, 355 (1993).

5. D. Chakrabarty, T. Koh, L. Bildsten, \etal, Astrophys.
    J. 446, 826 (1995).

6. D. K. Galloway, E. H. Morgan, and A. M. Levine,
    Astrophys. J., (2004) (in press); astro-ph/0401476.

 7. P. Ghosh and F. Lamb, Astrophys. J. 234, 296 (1979).

 8. V. M. Kaspi and F. P. Gavriil, Astrophys. J. 596, L71
    (2003).

 9. A. Krawczyk, A. G. Lyne, J. A. Gil, \etal, MNRAS
    340, 1087 (2003).

 10. F. Lebrun, J. P. Leray, P. Lavocat, \etal, Astron.
    Astrophys. 411, L141 (2003).

 11. X.-D. Li and Z.-R. Wang, Astron. Astrophys. 307, L5
    (1996).

12. V. M. Lipunov, Astrophysics of Neutron Stars (Nau-
    ka, Moscow, 1987)[in Russian].

13. N. Lund, S. Brandt, C. Budtz-Joergesen, \etal, As-
    tron. Astrophys. 411, L231 (2003).

14. A. A. Lutovinov, S. A. Grebenev, R. A. Sunyaev,
    \etal, Astron. Lett. 20, 538 (1994) [Pis'ma Astron. 
    Zh. 20, 631 (1994)].

15. A. A. Lutovinov, S. V. Molkov and M. G. Revnivstev,
    Astron. Lett. 29, 713 (2003) [Pis'ma Astron. Zh. 29, 
    803 (2003)].

16. F. Nagase, Publ. Astron. Soc. Japan 41, 1 (1989).

17. I. Negueruela, G. L. Israel, A. Marco, \etal, Astron.
    Astrophys. 397, 739 (2003).

18. A. N. Parmar, N. E. White, and L. Stella, Astrophys.
    J. 338, 373 (1989).

19. M. G. Revnivstev, R. A. Sunyaev, D. A. Varshalovich,
    \etal, Astron. Lett. 30, 382 (2004) [Pis'ma Astron. 
    Zh. 30, 430 (2004)].

20. J. Swank and E. Morgan, IAU Circ. No. 7531 (2000).

21. P. Ubertini, F. Lebrun, G. Di Cocco, \etal, Astron.
    Astrophys. 411, L131 (2003).

22. N. White, J. Swank, and S. Holt, Astrophys. J. 270,
    771 (1983).

23. C. Winkler, T. J.-L. Courvoisier, G. Di Cocco, \etal,
    Astron. Astrophys. 411, L1 (2003).

\pagebreak

\begin{table}[h]
\centering
{\bf Table 1.}  IBIS observationsa of the pulsar KS 1947+300 \,\a

\vspace{2mm}

\begin{tabular}{c|c|c|c} \hline \hline
Time of observations, MJD&Orbital phase&Flux, mCrab&Period, s  \\ 
\hline
&&&\\ [-4mm]
52605--52615 &0.33--0.58& 4.5$\pm$0.3 & 18.669$\pm$0.001\\[1mm]
52723 &0.25& 6.9$\pm$2.1 & --\b \\[1mm]
52746 &0.82& 12.1$\pm$2.3 & --\b\\[1mm]
52761 &0.19& 64.0$\pm$2.8 & 18.721$\pm$0.005 \\[1mm]
52770 &0.42& 77.8$\pm$2.7 & 18.718$\pm$0.002 \\[1mm]
52782 &0.71& 67.8$\pm$2.3 & 18.721$\pm$0.002 \\[1mm]
52797 &0.08& 15.3$\pm$2.0 & --\b \\[1mm]
52806 &0.31& 6.3$\pm$2.5 & --\b \\[1mm]
52821 &0.68& 7.7$\pm$2.3 & --\b \\[1mm]
52970 &0.36& 31.0$\pm$13.9 & --\b \\[1mm]
52985 &0.74& 69.8$\pm$3.4 & 18.727$\pm$0.002 \\[1mm]
52994 &0.96& 47.5$\pm$2.4 & 18.725$\pm$0.002 \\[1mm]
53019 &0.58& 6.3$\pm$2.1 & --\b\\[1mm]
53102 &0.65& 96.1$\pm$2.9 & 18.730$\pm$0.002 \\[1mm]
\hline
\end{tabular}

\vspace{3mm}

\begin{tabular}{cl}

\a& In the energy range 18-60 keV.\\
\b& Cannot be determined due to the poor statistics.\\

\end{tabular}
\end{table}

\pagebreak

\begin{landscape}
\begin{table}[h]
\centering

\hspace{-3mm}{\bf Table 2.}{ Best-fit parameters for the spectrum of KS 1947+300\a}

\vspace{2mm}

\hspace{-5mm}\begin{tabular}{@{}c|c|c|c|c|c}
\hline\hline
&&&&\\ [-4mm]
 Date &  $\alpha$ &$E_{Fe}$\b, keV& $E_{c}$, keV & $E_{f}$, keV
 &$\chi^{2}_{\,N} (N)$\c \\[2mm]

\hline
\multicolumn{6}{c}{}\\[-4mm]
\multicolumn{6}{c}{Based on INTEGRAL (IBIS) data}\\[2mm]
\hline
&&&&&\\ [-4mm]
MJD 52605--52630 &$1.07$\d&--&$8.6$\d&$26.4\pm2.7$&2.91(7)\\[2mm]
MJD 52770 &$1.07$\d&--&$8.6$\d&$24.7\pm1.9$&2.30(7)\\[2mm]
MJD 52985 &$1.07$\d&--&$8.6$\d&$26.5\pm3.4$&0.66(7)\\[2mm]
\hline
\multicolumn{6}{c}{}\\[-4mm]
\multicolumn{6}{c}{Based on INTEGRAL (JEM-X + IBIS) data}\\[2mm]
\hline
&&&&&\\ [-4mm]
MJD 53102 &$1.07^{+0.24}_{-0.13}$&--&$8.6^{+3.4}_{-1.2}$&$23.6^{+5.3}_{-2.3}$&1.18(104)\\[2mm]
\hline
\multicolumn{6}{c}{}\\[-4mm]
\multicolumn{6}{c}{Based on RXTE (PCA + HEXTE) data}\\[2mm]
\hline
&&&&&\\ [-4mm]
February 2001&$1.38\pm0.01$&$6.47\pm0.07$&$15.8\pm0.5$&$34.2\pm0.7$&1.33(98)\\[2mm]
March 2001&$1.26\pm0.01$&$6.55\pm0.08$&$12.5\pm0.5$&$28.5\pm0.4$&1.34(112)\\[2mm]
April 2001&$1.13\pm0.02$&$6.34\pm0.09$&$11.4\pm0.4$&$25.4\pm0.7$&0.99(117)\\[2mm]
May 2001 &$0.82\pm0.08$&$6.65\pm0.11$&$6.5\pm0.5$&$18.5\pm1.4$&1.03(117)\\[2mm]
June 2001\e &$0.84\pm0.24$&--&$5.7\pm1.2$&$11.6\pm3.3$&0.63(35)\\[2mm]
\hline
\end{tabular}

\vspace{3mm}

\begin{tabular}{cl}
\a& All errors are given at the 1$\sigma$ level.\\
\b& The position of the line center.\\
\c& The $\chi2$ value normalized to the number of degrees of freedom N.\\
\d& The parameters are fixed.\\
\e& Based only on PCA (3-20 keV) data.\\

\end{tabular}
\end{table}
\end{landscape}

\pagebreak

\begin{figure*}[t]
\includegraphics[width=6cm,bb=40 220 230 535]{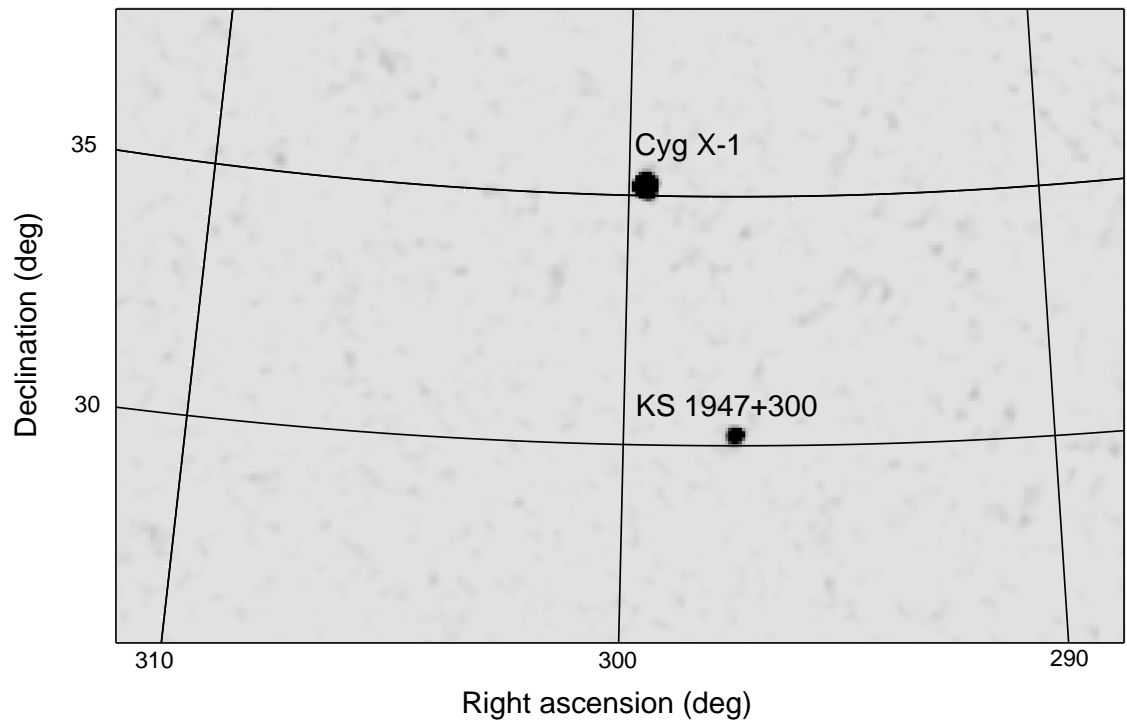}

\vfill

\caption{Sky map with KS 1947+300 obtained by 
the IBIS telescope in the energy range 18-60 keV. 
The total exposure time
was about 6 ks.
  }
\end{figure*}

\newpage

\begin{figure*}[t]
\centerline{\includegraphics[width=18cm,bb=18 190 570 690]{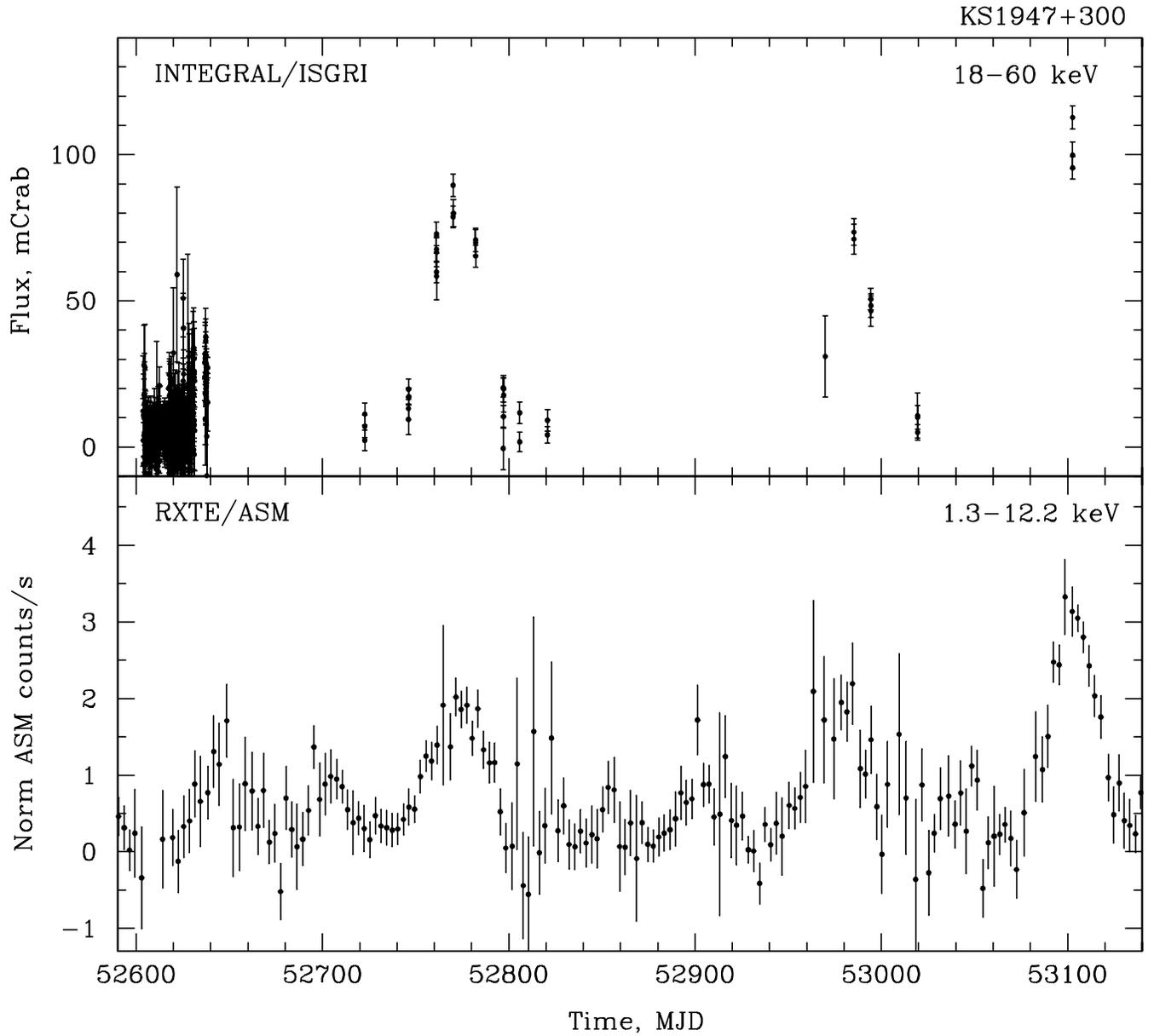}}

\vfill

\caption{Light curves for the pulsar KS 1947+300 in 
the energy ranges (a) 18-60 and (b) 1.3-12.2 keV. 
The errors correspond
to one standard deviation.  }
\end{figure*}

\newpage

\begin{figure*}[t]
\includegraphics[width=14cm,bb=70 260 500 705,clip]{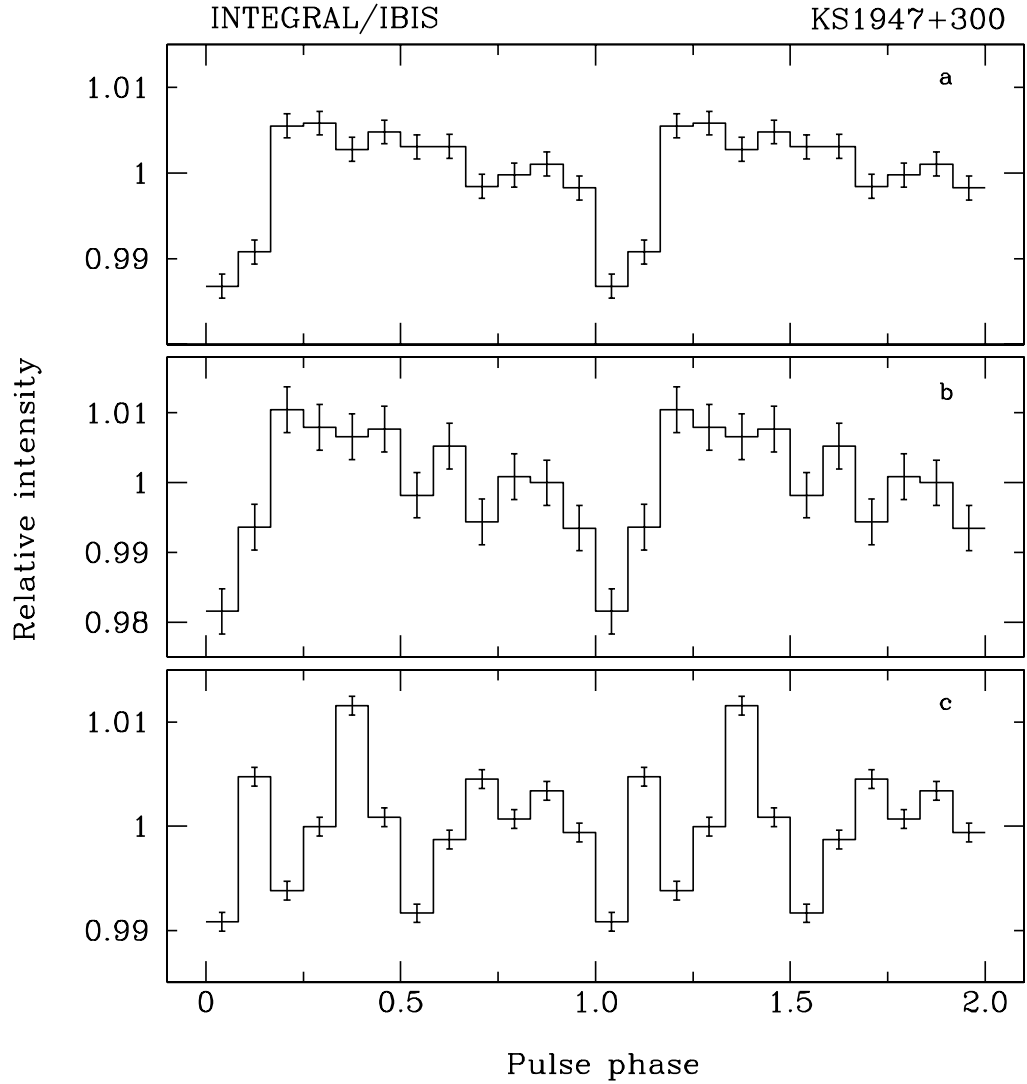}

\vfill

\caption{IBIS pulse profiles for KS 1947+300 at differ-
ent intensities; the mean bolometric luminosity of the
source is (a) $2.5\times10^{37}$ (MJD 52770), 
(b) $1.5\times10^{37}$ (MJD 52994), and (c) $0.2\times10^{37}$ 
erg s$^{-1}$ (MJD 52605-52615).  }
\end{figure*}

\newpage

\begin{figure*}[t]
\includegraphics[width=14cm,bb=70 145 495 715,clip]{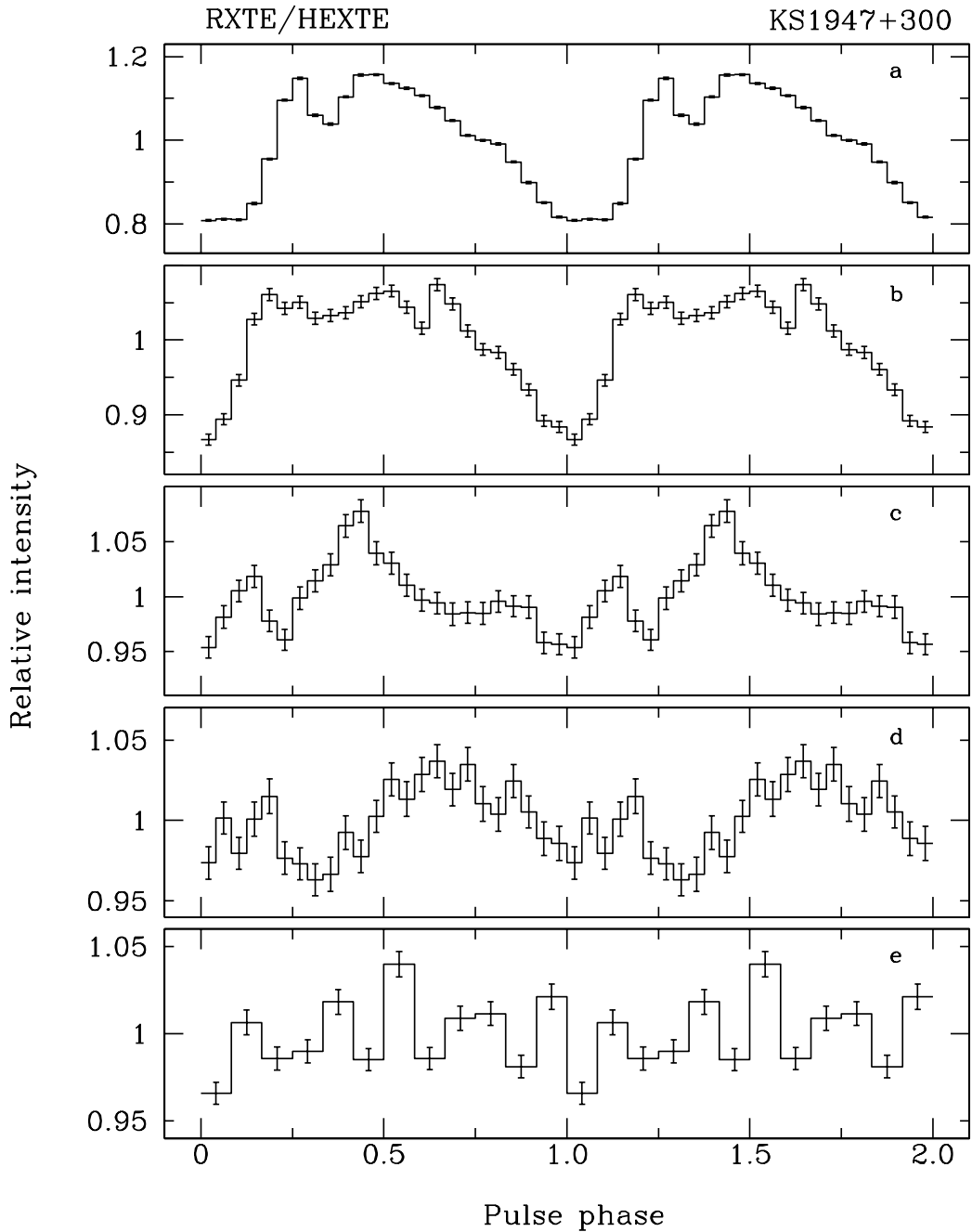}

\vfill

\caption{HEXTE pulse profiles for KS 1947+300 at 
different intensities during the outburst of 2000-2001 (the
background was not subtracted). The mean bolometric
luminosity of the source is 
$10.6\times10^{37}$ (a), $5.4\times10^{37}$ (b), 
$3.4\times10^{37}$ (c), $0.9\times10^{37}$ (d), 
$0.3\times10^{37}$ (e) erg s$^{-1}$. }
\end{figure*}

\newpage

\begin{figure*}[t]
\includegraphics[width=14cm,bb=75 260 500 705]{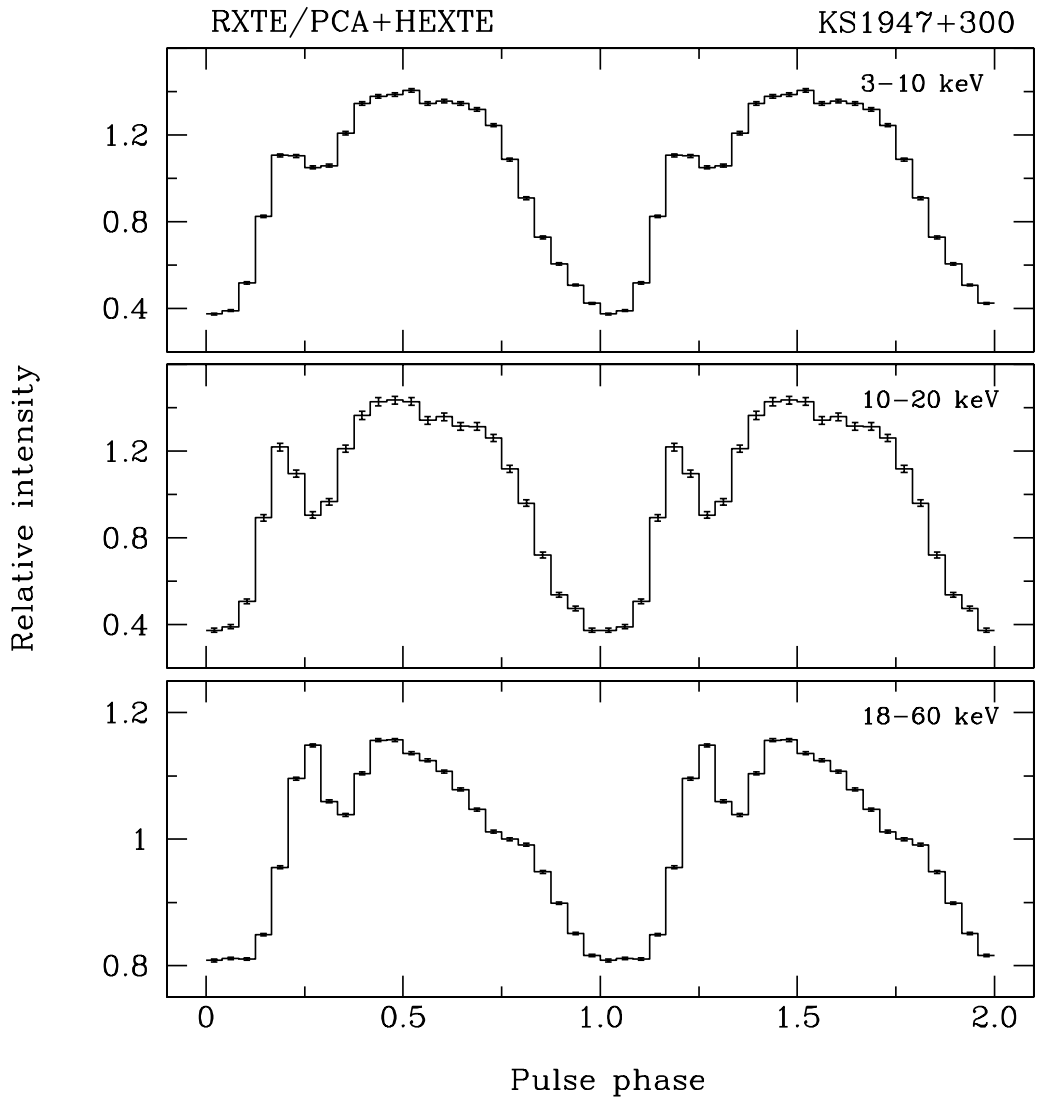}

\vfill

\caption{RXTE pulse profiles for KS 1947+300 in different
energy ranges. The errors correspond to one standard
deviation. }
\end{figure*}

\newpage

\begin{figure*}[t]
\includegraphics[width=14cm,bb=75 180 500 710,clip]{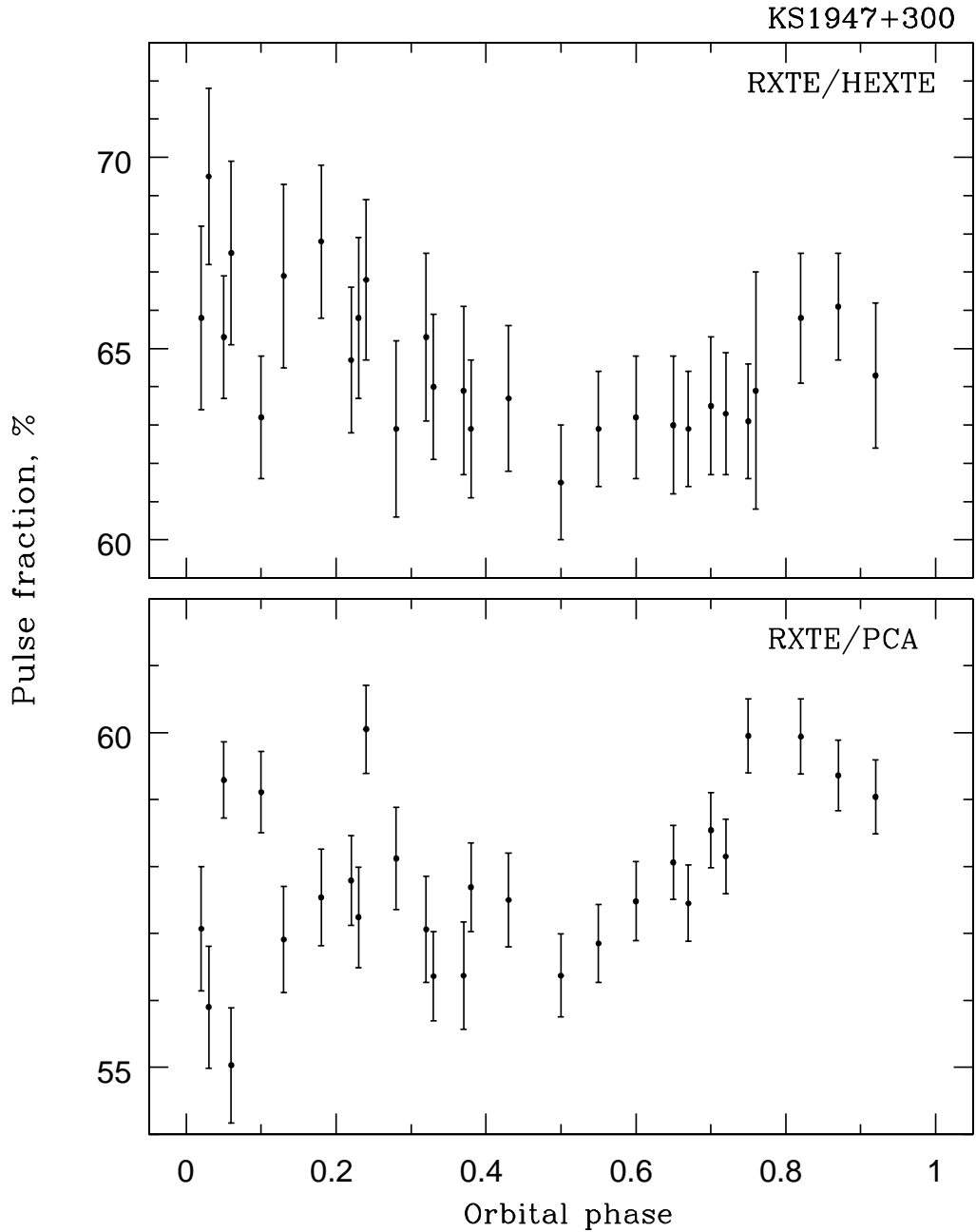}

\vfill

\caption{Pulse fraction in the energy range 18-60 keV
versus orbital phase of the pulsar KS 1947+300 during
the 2000-2001 outburst, as derived from the HEXTE/RXTE 
data (a), and in the energy range 3-20 keV, as
derived from the PCA/RXTE data (b). }
\end{figure*}

\newpage

\begin{figure*}[t]
\includegraphics[width=14cm,bb=95 240 515 690]{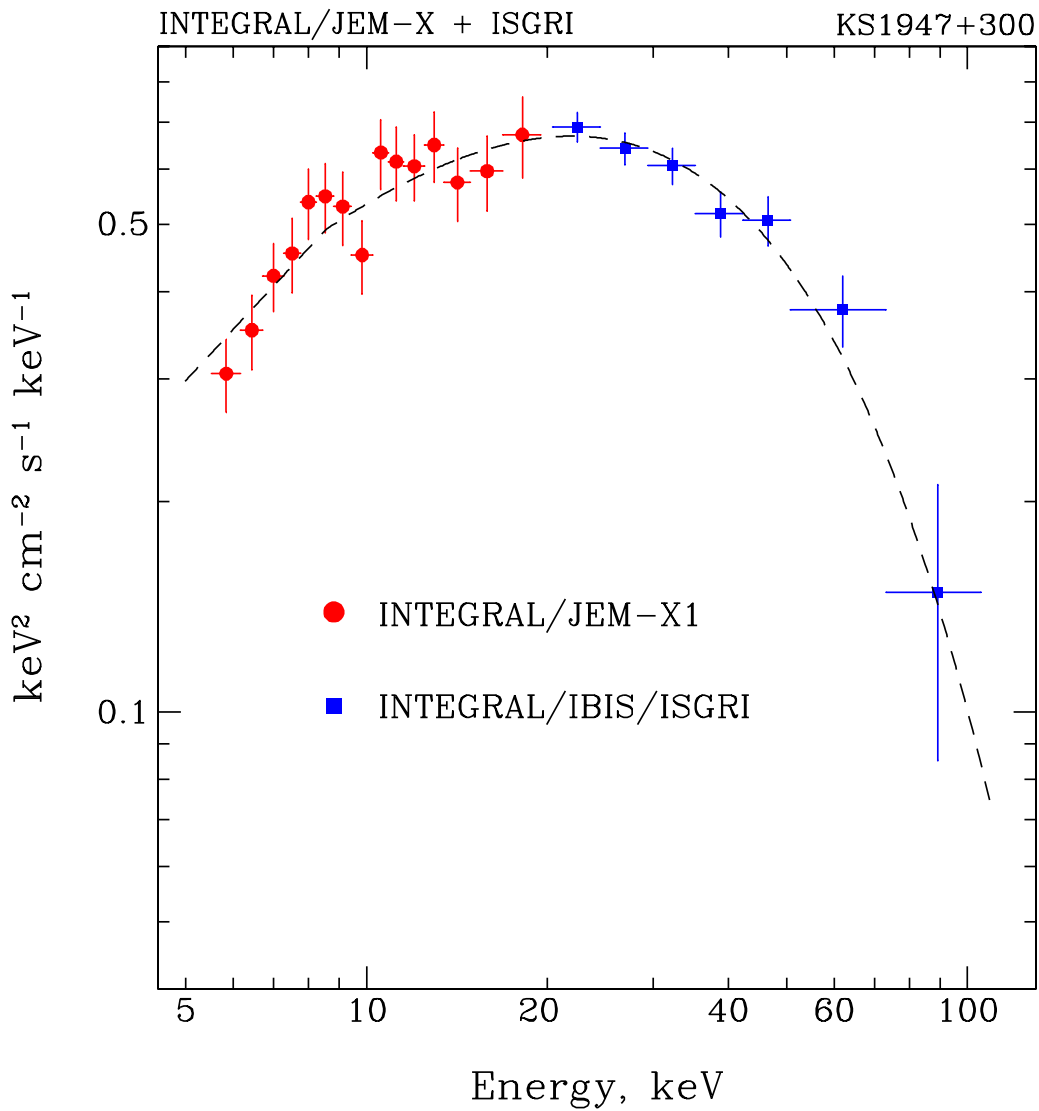}

\vfill

\caption{Energy spectrum for KS 1947+300, as 
constructed from the observations by the JEM-X and IBIS
telescopes aboard INTEGRAL on April 7, 2004. The
dashed line indicates the best fit to the spectrum. }
\end{figure*}

\clearpage

\begin{figure*}[t]
\includegraphics[width=14cm,bb=95 240 515 690]{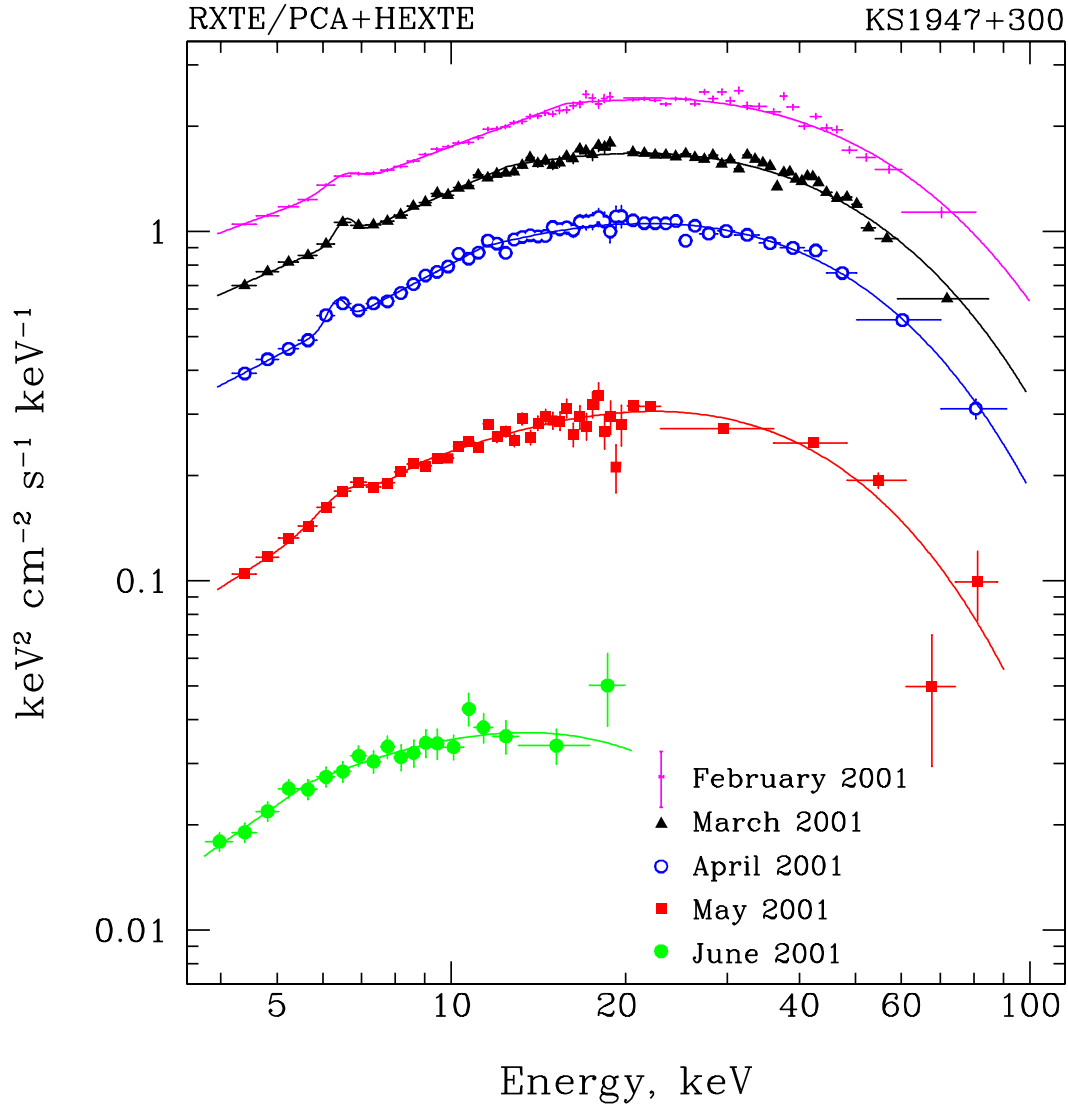}

\vfill

\caption{Energy spectra for KS 1947+300 at different luminosities during the 2000-2001 outburst, as constructed from the
RXTE data. The solid lines represent a power-law fit to the spectrum with a high-energy cutoff.}
\end{figure*}

\clearpage

\begin{figure*}[t]
\includegraphics[width=14cm,bb=55 305 500 705,clip]{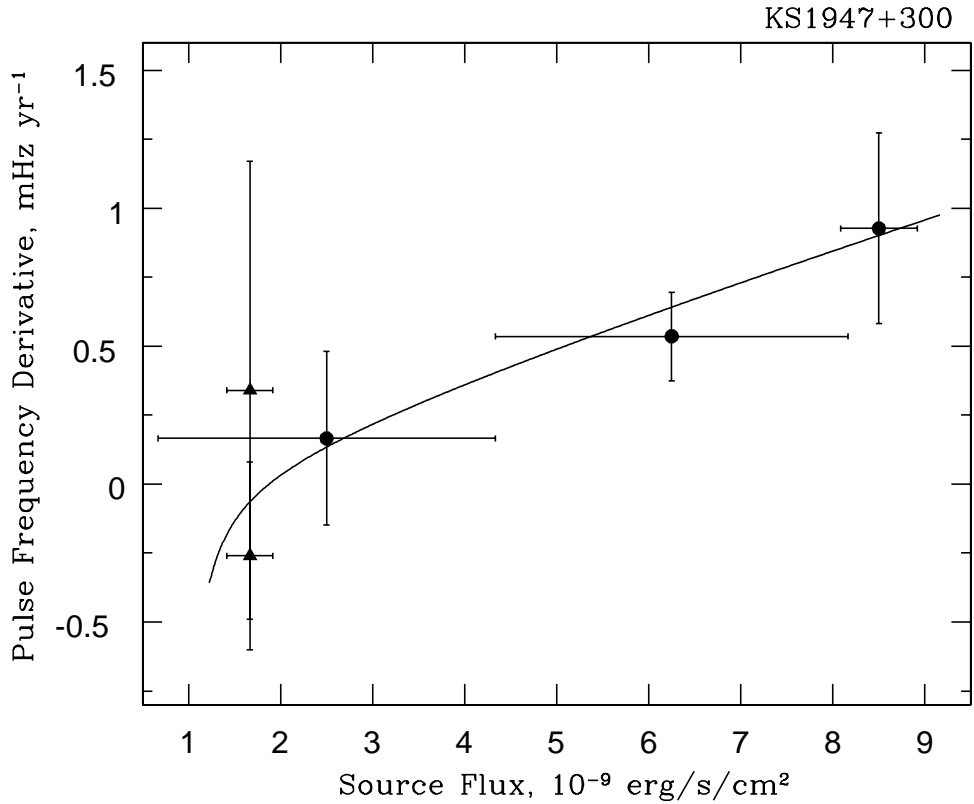}

\vfill

\caption{Luminosity dependence of the rate of change in
the pulsation frequency of KS 1947+300, as derived from
the INTEGRAL (triangles) and RXTE (circles) data. The
solid line indicates the model dependence for a distance
to the binary of $d\sim9.5$ kpc and a magnetic field of $B\sim 
2.5\times10^{13}$ G.  }
\end{figure*}

\end{document}